# A new model for the double well potential

Guihua Tian[1,*]

[1]*School of Science, Beijing University of Posts And Telecommunications.
Beijing 100875, China.*
(Dated: Jan 1, 2010)

A new model for the double well potential is presented in the paper. In the new potential, the exchanging rate could be easily calculated by the perturbation method in supersymmetric quantum mechanics. It gives good results whether the barrier is high or sallow. The new model have many merits and may be used in the double well problem.

**Introduction**

The quantum tunneling in double well potential have been extensively applied in many branches of physics. In spite of being almost as old as quantum mechanics, the tunneling dynamics of the double well are one of the most active research areas. For example, they appeared in the mean-fields dynamics of Bose-Einstein condensates, the recent development of ion trap technology, the ultracold trapped atoms theory and applications, etc[1]-[5]. In all these areas, the single particle energy splitting, the energies and the wave functions in the double well are needed for the relevance problem. the usual potential model of the double well for the single particle is

$$V(x) = -ax^2 + bx^4. \qquad (1)$$

The model gives rise to the challenge to the mathematical and computational techniques. The focus of the problem of finding the energy splitting due to the tunneling turned out to be difficult. It can not be treated by the regular perturbation methods. Then many methods appeared to attack the problem, like the instanton method, the WKB approximation method, Variational method, numerical method, and the improved WKB method, the super-symmetry quantum method, etc. The validity of these methods are still limited.

It is the form of the potential that make all the efforts of the above mentioned methods not satisfactory. Because the property of the potential (1) at the infinity $x = \infty$, the singular point of the Schröedinger's, changes steeply from $b = 0$ to $b \neq 0$, it presents one of the greatest challenge to the methods of solving differential equations. Actually, no matter how small the parameter $b$ might be, the perturbation term $bx^4$ would absolutely exceeds the un-perturbation term $ax^2$ when $x$ is large enough. So the property at the infinity of the potential and the Schrödinger equation is mainly determined by the term $bx^4$, which is no longer the small perturbation term. Hence the regular perturbation method for small $b$ to solve the double problem does not work for sure. It is natural that there still have some controversial conclusions on how reliable the above mentioned methods may be.

As stated before, it is the potential form of (1) that gives rise to the difficulties shown in treating it. The best way to the problem is to have some models other than that of the (1). The ideal model should be solved analytically. There is a model of the delta split harmonic oscillator almost meets the requirement [6]:

$$V(x) = \frac{1}{2}m\omega^2 x^2 + \kappa\delta(x) \qquad (2)$$

Though there are some possible situations for the ideal or perfect point-like potential, it is somehow still a mathematical idealization.

Hence another solution that naturally comes forth is to give a new physical model. The new model should have the traits of the double potential. It better had completely solvable property, if not, it should allow the regular perturbation methods to apply. In the paper, we propose a physical model meeting all these requirements for the double well potential. Actually, we have found it in studying the spheroidal equations.

In solving spheroidal wave functions, we have connected them with the following equation

$$\frac{d^2\Psi}{dx^2} + \left[\frac{1}{4} + \alpha\cos^2 x - \frac{m^2 - \frac{1}{4}}{\sin^2 x} + E\right]\Psi = 0 \qquad (3)$$

and the boundary conditions for $\Psi$ become

$$\Psi|_{x=0} = \Psi|_{x=\pi} = 0 \qquad (4)$$

The the potential $V(x, \alpha, m)$ is

$$V(x, \alpha, m) = -\frac{1}{4} - \alpha\cos^2 x + \frac{m^2 - \frac{1}{4}}{\sin^2 x} \qquad (5)$$

where the parameters $m$ are integers. The eqn.(3) have been solved by the perturbation methods in supersymmetry quantum mechanics[9],[10]. See appendix for the detail of the complete solution.

The potential is symmetry with the center at $x = \frac{\pi}{2}$. It has many interesting properties. When the parameter $m = 0$, the potential $V(x, \alpha, 0)$ behaves as the figures shown for the ranges $\alpha > -\frac{1}{4}$ and $\alpha < -\frac{1}{4}$ respectively. Only in the case $\alpha < -\frac{1}{4}$, does it possess two local minimums and look like a reverse double well potential.

---

*Electronic address: tgh-2000@263.net,tgh20080827@gmail.com

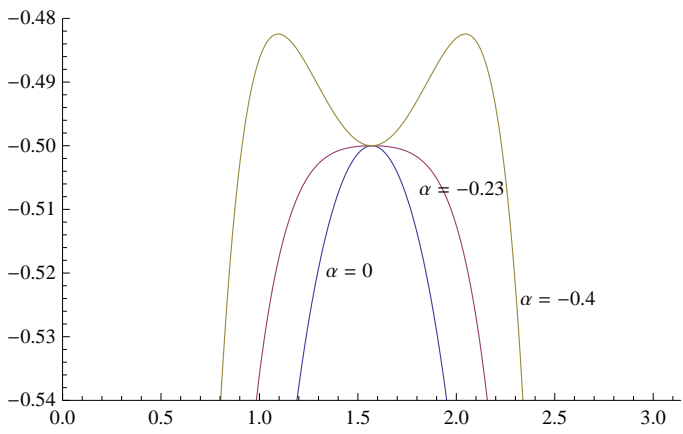

FIG. 1: comparison of the potentials $V(x,\alpha,0)$ in the case $m=0$. The blue, red and yellow lines are the potential with $\alpha=0,\ -0.23,\ -0.4$. The figure shows the potential changes its properties when $\alpha < -\frac{1}{4}$.

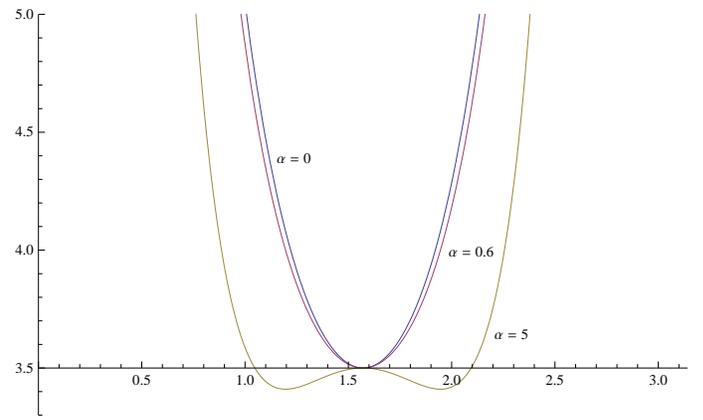

FIG. 2: comparison of the potentials $V(x,\alpha,2)$ in the case $m=2$. The blue, red and yellow lines are the potential with $\alpha = 0,\ 0.6,\ 5$. The figure shows the formation of the double well potential when $\alpha > 4 - \frac{1}{4}$.

When $m \neq 0$, the potential changes greatly, and approaches positive infinity at $x = 0,\ \pi$. In the range $\alpha < m^2 - \frac{1}{4}$, the difference of the potentials between $\alpha = 0$ and $\alpha \neq 0$ is artificial. The fundamental changes come when $m^2 - \frac{1}{4} < \alpha$. The nice property of the potential comes for $m^2 - \frac{1}{4} < \alpha$: it resembles that of the form of the double potential. The potential has two equal minimum values

$$V_{min} = 2\sqrt{\alpha(m^2 - \frac{1}{4})} - \frac{1}{4}\alpha \qquad (6)$$

at the symmetrical points $x_{min,1}, x_{min,2}$ satisfying

$$\sin^4 x_{min,1} = \sin^4 x_{min,2} = \frac{m^2 - \frac{1}{4}}{\alpha}. \qquad (7)$$

The maximal value is achieved at the center point $x = \frac{\pi}{2}$,

$$V_{max} = m^2 - \frac{1}{2} \qquad (8)$$

the barrier's height is

$$V_{max} - V_{min} = \left[\sqrt{\alpha} - \sqrt{m^2 - \frac{1}{4}}\right]^2 \qquad (9)$$

It is easy to see that the depth of the double well is completely determined by the parameter $\alpha$ when the integral quantity $m$ is fixed. The depth will increase as $\alpha$ goes up. Small $\alpha$ corresponds the sallow double well.

The great advantage of the potential (5) lies in that the properties of the singularities of the potential do not be influenced by the term $-\alpha \cos^2 x$, which only adds a constant $-\alpha$ to the infinity at $x = 0$ or $x = \pi$. This stands in contrast with that in the potential (1). The term $-\alpha \cos^2 x$ actually brings the double-well-form potential into reality. This is just what we want to achieve. Hence, one can treat the problem by the usual perturbation methods, which have been studied in series papers[7]-[11].

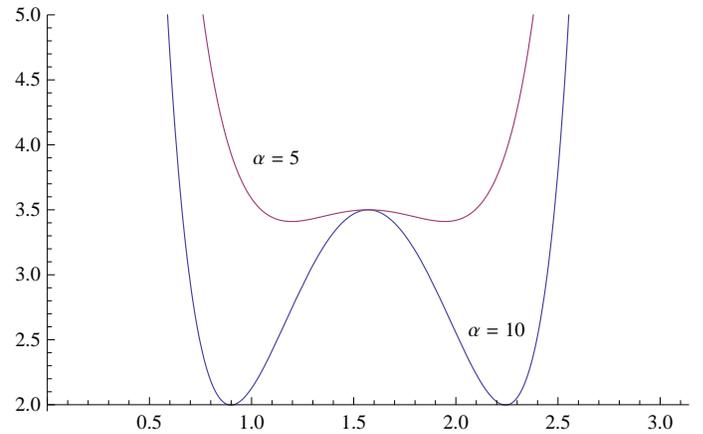

FIG. 3: comparison of the potentials $V(x,\alpha,0)$ in the case $m = 2$. The blue, red lines are the potential with $\alpha = 10,\ 5$. The figure shows the double well depth decreases as $\alpha$ decreases.

In the double well problem, the tunneling rate is the energy difference $t = E_1 - E_0$ between the lowest wave functions. The focus is how accurate one can get the tunneling rate through the double barrier. In the new model, this corresponds to the eigen-energy difference of $t = E_{1;m} - E_{0;m}$, which has been given in previous works

[9], [10] as

$$E_{1;m}^{-} - E_{0;m} = R(a_1), \ R(a_1) = \sum_{l=0}^{\infty} \alpha^l R_l(a_1), \quad (10)$$

$$a_1 = (A_{i,j}), \ A_{i,j} = 1, \quad (11)$$
$$a_2 = (B_{i,j}), \ B_{i,j} = B_{i,j}(A_{l,k}) \quad (12)$$
$$R_0 = (2m+2) \quad (13)$$
$$R_1 = -\frac{B_{1,1} + A_{1,1}}{2m+3} \quad (14)$$
$$R_n(A_{i,j}) = \left[c_{n,1}^{+}(A_{n,j}) - c_{n,1}^{-}(B_{n,j})\right], n \geq 2 \quad (15)$$

where

$$c_{n,1}^{+}(A_{n,1}) = -\left[(2m+1)A_{0,0} - 1\right]A_{n,1}a_{n,1} \quad (16)$$
$$c_{n,1}^{-}(B_{n,j}) = -\left[(2m+1)B_{0,0} + 1\right]B_{n,1}a_{n,1} \quad (17)$$

See the reference [10] for the determination of the quantities $a_{i,j}, B_{i,j}$. Here we give some example, when $m = 2$,

$$R_0 = 12, \ R_1 = -0.19047619, \quad (18)$$
$$R_2 = -0.001374287. \quad (19)$$

From the relation (10), we have

$$\alpha = 5 \to t = E_1 - E_0 = 11.01326187, \quad (20)$$
$$\alpha = 10 \to t = 9.9578094 \quad (21)$$

respectively. The potential (5) represents the symmetry double potential, and the asymmetry potential could be represented by

$$V(x, \beta, m, s) = \frac{(m + s\cos x)^2 - \frac{1}{4}}{\sin^2 x} - \beta^2 \cos^2 x$$
$$+ 2s\beta \cos x - \frac{1}{4} - s. \quad (22)$$

This potential appears in the study of the spin-weighted spheroidal functions[15], [16],[17]. The spin-weighted spheroidal functions come from the study of the perturbation problem of the Kerr black hole[12]-[14]. The similar results about the eigen-values and eigen-functions to that of the references [9],[10] will be reported in the future. This potential approaches infinity at $x = 0, \pi$ and might have two minimums and one maximum in some range of the parameter $\alpha$. The graph of the potential is similar to that of the asymmetry double well in some way: their distances to the center $x = \frac{\pi}{2}$ not equal. However, the two depthes of the asymmetry potential are not equal.

The double well potential (1) have been applied extensively in physics. Their fundamental properties are in abundance depending the parameters $a$, $b$. In the paper, a new type of potential (5) is proposed. They are shown to have the similar fundamental properties as that for the double well ones (1). The nice properties of the new potential is that their Schrödinger equations could be solved by the usual perturbations methods, which stands in vivid contrast against the one (1). Furthermore, their solutions in perturbation series of the parameter $\alpha$ are already given in the [9],[10]. All these results are useful for their application in the double well problem.

**Acknowledgements**

This work was supported in part by the National Science Foundation of China under grant Nos.10875018, 10773002, 2010CB923200.